\documentclass[10pt,article]{article}
\usepackage[top=0.85in,left=0.75in,footskip=0.75in,marginparwidth=2in]{geometry}
\usepackage{physics}
\usepackage{nameref,hyperref}
\usepackage{graphicx}
\usepackage{color}
\usepackage{abstract,lipsum}
\usepackage{cite}

\usepackage[aboveskip=1pt,labelfont=bf,labelsep=period,singlelinecheck=off]{caption}

\makeatletter
\renewcommand{\@biblabel}[1]{\quad#1.}
\makeatother

\usepackage{lastpage,fancyhdr,graphicx}
\usepackage{epstopdf}
\fancyheadoffset[L]{2.25in}
\fancyfootoffset[L]{2.25in}

\usepackage{wrapfig}
\usepackage[pscoord]{eso-pic}
\usepackage[fulladjust]{marginnote}
\reversemarginpar

\begin{document}
\vspace*{0.35in}

\begin{flushleft}
{\LARGE
\textbf\newline{ Trap split with Laguerre-Gaussian beams}
}
\newline
\\
Seyedeh Hamideh Kazemi\textsuperscript{1},
Saeed Ghanbari\textsuperscript{1},
Mohammad Mahmoudi\textsuperscript{1,*}
\\
\bigskip
\textsf{1} Department of Physics, University of Zanjan, University Blvd., 45371-38791, Zanjan, Iran
\\
* mahmoudi@znu.ac.ir

\end{flushleft}
\begin{abstract}
We present a convenient and effective way to generate a novel phenomenon of trapping, named trap split, in a conventional four-level double-$\Lambda$ atomic system driven by four femtosecond Laguerre-Gaussian laser pulses. We find that trap split can be always achieved when atoms are trapped by such laser pulses, as compared to Gaussian ones. This feature is enabled by interaction of the atomic system and the Laguerre-Gaussian laser pulses with zero intensity in the center. A further advantage of using Laguerre-Gaussian laser pulses is the insensitivity to fluctuation in intensity of the lasers in such a way that the separation between traps remains constant. Moreover, it is demonstrated the suggested scheme with Laguerre-Gaussian laser pulses can form optical traps with spatial sizes that are not limited by the wavelength of the laser and can, in principle, become smaller than the wavelength of light. This work would greatly facilitate the trapping and manipulating the particles and generation of trap split. It may also suggest the possibility of extension into new research fields, such as micro-machining and biophysics.
\end{abstract}
\vspace*{0.2in}



 Since Ashkin's seminal work on trapping a particle through the use of radiation forces exerted by a Gaussian laser beam \cite{Ashkin1}, the optical trapping techniques have been extensively used in physics, biophysics, micro-chemistry, and micro-mechanics to allow trapping and manipulation of neutral atoms, nanoparticles, cells, biological substances, DNA and RNA molecules as well as microsized dielectric particles \cite{block,1,2,3,4,5,6,jiang,9,kemp}. However, as these techniques are based on the optical gradient force which is dependent on the size of the particle \cite{Ashkin}, smaller particles are much more difficult to trap \cite{Li,Tsuboi}. Enhancement of the gradient force by using the femtosecond (fs) pulses could provide an efficient method to address the issue. During recent years, due to the tremendous technological advancement in the generation of fs lasers, there has been a resurgence of interest in this field \cite{liu,Baylam,liu2} and the optical trapping technologies have been further developed \cite{agate,pan,sanz,11,usman}. Tamai \textit{et al.} reported the feasibility of achieving stable laser trapping of water-soluble CdTe quantum dots by using a high repetition-rate picosecond Nd:YLF laser with two orders of magnitude lower than that used in continuous wave laser trapping \cite{pan}. Not only does fs laser lead to efficient trapping \cite{pan,agate}, but also it results in discoveries of novel physical phenomena including the deposition of nanostructured CdS films \cite{sanz} and controlling the direction of the scattered nanoparticles \cite{usman}.

 Recently, Okamoto's group observed the stable trapping of 60-nm gold nanoparticles when they are trapped by the fs laser pulses. They have discovered a new trapping phenomenon due to the nonlinear optical effect in which the stable trap site splits into two positions, as the incident peak-laser power approaches a threshold level, named trap split \cite{jiang}. More recently, Chakraborty and Sarma have shown how to control optical trap potential (OTP) in an N-type four level atomic system by tuning the beam waist of the chirped fs Gaussian pulses and the detuning frequency \cite{kumar5}. They have shown that the potential splits with increasing the Rabi frequency, a behavior analogous to that in trapping of nonparticles with fs pulses observed in Okamoto's study. However, this method is not proper for a variety of tasks, especially for biological samples in which increasing of the laser radiation intensity may result in damage. In a recent paper by two authors of the current paper (S. H. Kazemi and M. Mahmoudi), OTP was studied in a four-level double-$\Lambda$ closed-loop atomic system in multiphoton resonance condition and, more specifically, they found that the OTP splits into two positions, through the special switching of the relative phase of applied fields \cite{kazemi}. 
It may be noted that, there has been some work on the Bose gases in a split trap. In 2003, Busch and Huyet investigated low-density, one-dimensional quantum gases in a split trap. By calculating many-body wave-functions, it was shown that by increasing strength of the splitting potential, a central notch in ground-state density gets deeper \cite{busch}. To our knowledge, none of the previous studies on trap split have provided the possibility to permanently generate trap split, although the various parameters have been used to control this emerging behaviour, such as initial conditions of atoms, relative phase, strength of potential, frequency and intensity of the applied fields. As this phenomenon in the previous works has occurred for limited set of those parameters, the results dramatically emphasize the needs for alternative or additional parameters to readily generate trap split.

 On the other hand, much of the previous literature on the topic has assumed plane-wave or Gaussian modes, although the use of Laguerre-Gaussian (LG) light beams \cite{allen,he}, having a doughnut-shaped intensity distribution and zero intensity at the beam center, induced new interest \cite{wright,aksenov,chen2,Kazemi2}, especially in optical trapping \cite{otsu}. For example, LG beams have given birth to various excellent applications such as rotating trapped microscopic particles \cite{Paterson} and singular optical lattice generation \cite{soar}. \\~\\
In the present paper, we investigate the OTP in a four-level double-$\Lambda$ type system by using fs $LG_{0}^{1}$ pulses and exploit the fact that the trap split always occurs using the LG pulses which may hold great promise for practical applications in trapping and manipulating the particles. We must reiterate the importance of the fact that trap split can be always achieved, while above studies involving the trapping of atoms by fs laser pulses, mainly concentrated on controlling this phenomenon.

\label{sec:examples}
Following the Ref. \cite{kazemi}, where a scheme to control the OTP in a closed-loop atomic system was proposed, we consider a four-level double-$\Lambda$ type system interacting with four coherent few-cycle pulse laser fields, whose level structure consists of two metastable lower states $\vert 1\rangle$ and $\vert 2\rangle$ plus two excited states $\vert 3\rangle$ and $\vert 4\rangle$. It should be noted that this model is the scheme that allows for a closed laser-field interaction loop in which optical properties of the medium could be controlled by the phases of the laser fields \cite{23}. All of the allowed electric dipole transitions, i.e., $ \vert 1\rangle - \vert 3\rangle$, $\vert 2\rangle - \vert 3\rangle$, $\vert 2\rangle - \vert 4\rangle$ and $\vert 1\rangle - \vert 4\rangle$ are driven by the laser fields.

 The electric field of the collimated linearly polarized laser beam interacting between $\vert i \rangle$ and $\vert j \rangle$ is defined as
\begin{equation}
\vec{E}_{ij} (r,t)= \vec{A}_{ij}(r,t) \cos(\omega_{ij} t +\omega_{D_{ij}}+\phi_{ij} ),
\label{eq1}
\end{equation}
where $\vec{A}_{ij}(r,t)$ is the space and time dependent field amplitude vector and $\omega_{ij}$ and $\phi_{ij}$ refer to the carrier frequency and the absolute phase of the pulse with the index $i, j$ being labelled four fields. Moreover, $\omega_{D_{ij}}= \vec{k_{ij}}. \vec{v}$ is the frequency detuning due to the translational motion with the wave vector of the corresponding electric field $\vec{k_{ij}}$.

 We now assume that the field amplitude has either a Gaussian profile or an $LG_{0}^{1}$ one; For the Gaussian profile, we have $\vec{A}_{ij}(r,t)= \hat{e}_{ij} \, E_{0G}\, \exp( -r^2/ w^2-t^2/ \tau^2)$ with $w$, $\tau$ and $\hat{e}_{ij}$ being beam waist, temporal width of the pulse and unit polarization vector, respectively. We also simplify the notation $E_{0ijG}$ to $E_{0G}$. Laguerre-Gaussian beam ($LG_{p}^{l}$) defines a solution of the paraxial wave function in a cylindrical coordinate which its indices $l$ and $p$ are the number of times the phase completes $ 2 \pi $ on a closed loop around the axis of propagation and the number of radial node for radius $r>0$, respectively \cite{allen}. In this paper, we use an $LG_{0}^{1}$ mode
\begin{equation}\label{eq2}
\vec{A}_{ij}=\hat{e}_{ij}\, E_{0LG}\, (\frac{ \sqrt{2}\, r\,e^{i\psi} }{w} )\, \exp( \frac{-r^2}{w^2}-\frac{t^2}{\tau^2} ),
\end{equation}
and we will present the results for incident nonvortex LG laser pulses with vanishing azimuthal number. It is worth pointing out that in order to compare the OTP by the LG and Gaussian pulses, it is necessary that the fields have the same total laser power, $P= \int f(r) \, 2 \pi r \ dr$ with $f(r)$ being the beam radius profile function. Throughout the paper, the field amplitudes are chosen in such a way that the fields have the same power.

The general expression for a Rabi frequency can be written as $g=(\vec{\mu}. \vec{E})/{\hbar}$, where $\vec{\mu}$ and $\vec{E}$ are the atomic dipole moment of the corresponding transition and peak amplitude of the field, respectively. The Rabi frequency for a Gaussian beam is $g=g_{0}\, \exp( -r^2/w^2-t^2/\tau^2)$ with $g_{0}=(\vec{E}_{0G}.\vec{\mu})/{\hbar}$. The corresponding expression for an $LG_{0}^{1}$ field is $g=(g^{'}_{0} \, \sqrt{2} \,r / w)$
$\exp( - r^2/ w^2-t^2/\tau^2) $ with $g^{'}_{0}=(\vec{E}_{0LG}.\vec{\mu})/{\hbar}$ as the Rabi frequency constant.

Under the dipole and rotating-wave approximations, the Hamiltonian in the interaction picture reads \cite{kazemi}

 \begin{eqnarray}\label{eq}
V &=&\hbar (\Delta_{32} -\Delta_{31}) \tilde{\rho}_{22}-\hbar \Delta_{31}\tilde{\rho}_{33} + \hbar (\Delta_{32} -\Delta_{31}-\Delta_{42}) \tilde{\rho}_{44}\\ \nonumber
&-&\hbar (g_{31}\tilde{\rho}_{31}+g_{32}\tilde{\rho}_{32}+g_{42}\tilde{\rho}_{42}+g_{41}\tilde{\rho}_{41} e^{-i\Phi}+H.c.).
\end{eqnarray}

 Here, $\rho_{lm} = \vert l \rangle \langle m \vert$, $\tilde{\rho}_{lm}\, (l,m \in \lbrace1, ..., 4\rbrace)$ and $\Delta_{ij}= \omega_{ij}-\bar{\omega}_{ij} $ denote the corresponding operator in the new reference frame and the detuning of the laser field, respectively. It is interesting to note that the residual time dependence in the system appears only together with the Rabi frequency $g_{41}$. Also, we define the relative phase of the applied fields as
\begin{equation}
\Phi = \Delta t- \vec{K} \vec{r} +\phi_{0},
\end{equation}

where the multiphoton resonance detuning, wave vector mismatch, and initial phase difference are given respectively by
\begin{subequations}
\begin{eqnarray}
\Delta &=& (\Delta_{32} +\Delta_{41})-(\Delta_{31} +\Delta_{42}), \\
\vec{K} &=& (\vec{k}_{32}+\vec{k}_{41})-(\vec{k}_{31}+\vec{k}_{42}), \\
\phi_{0} &=& (\phi_{32}+\phi_{41})-(\phi_{31}+\phi_{42}).
\end{eqnarray}
\end{subequations}
 From the Hamiltonian given by equation (\ref{eq}), the related density matrix equations for considered four-level system can be written as follows
\begin{subequations}
\begin{eqnarray}\label{eq66}
\dot{\tilde{\rho}}_{11} &=& i g_{31}^{*} \tilde{\rho}_{31}- i g_{31} \tilde{\rho}_{13}+ i g_{41}^{*} \tilde{\rho}_{41} e^{i\Phi} -i g_{41} \tilde{\rho}_{14} e^{-i \Phi}, \\
\dot{\tilde{\rho}}_{22} &=& i g_{32}^{*} \tilde{\rho}_{32}- i g_{32} \tilde{\rho}_{23}+ i g_{42}^{*} \tilde{\rho}_{42} - i g_{42} \tilde{\rho}_{24}, \\
\dot{\tilde{\rho}}_{33} &=&- i g_{31}^{*} \tilde{\rho}_{31}+ i g_{31} \tilde{\rho}_{13}- i g_{32}^{*} \tilde{\rho}_{32} + i g_{32} \tilde{\rho}_{23}, \\
\dot{\tilde{\rho}}_{12}&=& i ( \Delta_{32}-\Delta_{31} )\tilde{\rho}_{12} + i g_{31}^{*} \tilde{\rho}_{32}-i g_{32} \tilde{\rho}_{13}+i g_{41}^{*} \tilde{\rho}_{42}e^{i \Phi}-i g_{42} \tilde{\rho}_{14},\\
\dot{\tilde{\rho}}_{13}&=&- i \Delta_{31} \tilde{\rho}_{13} + i g_{31}^{*}( \tilde{\rho}_{33}- \tilde{\rho}_{11})- i g_{32}^{*} \tilde{\rho}_{12} +i g_{41}^{*} \tilde{\rho}_{43} e^{i\Phi},\\
\dot{\tilde{\rho}}_{14}\ &=& i(\Delta_{32}-\Delta_{31}-\Delta_{42}) \tilde{\rho}_{14}+ i g_{31}^{*} \tilde{\rho}_{34} - i g_{42}^{*} \tilde{\rho}_{12} \\ \nonumber
&+& i g_{41}^{*} e^{i\Phi} ( \tilde{\rho}_{44}- \tilde{\rho}_{11}),\\
\dot{\tilde{\rho}}_{23}\ &=&- i \Delta_{32} \tilde{\rho}_{23} + i g_{32}^{*}( \tilde{\rho}_{33}- \tilde{\rho}_{22})- i g_{31}^{*} \tilde{\rho}_{21} + i g_{42}^{*} \tilde{\rho}_{43},\\
\dot{\tilde{\rho}}_{24}\ &=&- i \Delta_{42} \tilde{\rho}_{24} + i g_{42}^{*}( \tilde{\rho}_{44}- \tilde{\rho}_{22})+i g_{32}^{*} \tilde{\rho}_{34} - i g_{41}^{*} \tilde{\rho}_{21} e^{i\Phi}, \\
\dot{\tilde{\rho}}_{34}\ &=&- i(\Delta_{42}-\Delta_{32}) \tilde{\rho}_{34} + i g_{31} \tilde{\rho}_{14} +i g_{32} \tilde{\rho}_{24} - i g_{42}^{*} \tilde{\rho}_{32}- i g_{41}^{*} \tilde{\rho}_{31} e^{i\Phi}.
\end{eqnarray}
\end{subequations}
In equations (6) the overdots stand for the time derivatives, and the remaining equations follow from the constraints $\tilde{\rho}_{lm}=\tilde{\rho}^{*}_{ml}$ and $\sum _{l} \tilde{\rho}_{ll}=1 $.

Generally, there are two kinds of interaction force between laser beam and atom, namely, reactive and dissipative forces. The first is due to the exchange of momentum between atoms and the various plane-wave composing the laser field, and the second is related to the absorption and emission of energy \cite{mystre, aksarma}. Here, due to the fast laser-atom interaction, we have only reactive force and the optical dipole force that we have driven is conservative. We can now use the Ehrenfest 's theorem \cite{cook} and derive the following expression for the transverse component of the optical dipole force and the OTP

 \begin{eqnarray}
F_{t} \ &=&\hbar [(\tilde{\rho}_{31}+\tilde{\rho}_{13}) \vec{\nabla} g_{31}+ (\tilde{\rho}_{32}+\tilde{\rho}_{23}) \vec{\nabla} g_{32} \\ \nonumber
&+& (\tilde{\rho}_{14}+\tilde{\rho}_{41}) \vec{\nabla} g_{41}+ (\tilde{\rho}_{24}+\tilde{\rho}_{42}) \vec{\nabla} g_{42}],\\ \nonumber
\vec{F}&=&-\vec{\nabla} U.
\label{eq6}
\end{eqnarray}

 Before we present the numerical results, it is desirable to point out some important considerations. As a realistic example, we consider hyperfine energy levels of $^{87}$ Rb \cite{steck}. It may be noted that as the scheme is independent of the chosen parameters, it can be easily implemented in current experimental settings. For simplicity, assume that the temporal width of each pulse is given by $\tau=20 \, fs$ and the Rabi frequencies are equal.
\begin{figure}[t]
\centering
\includegraphics[width=0.7\linewidth]{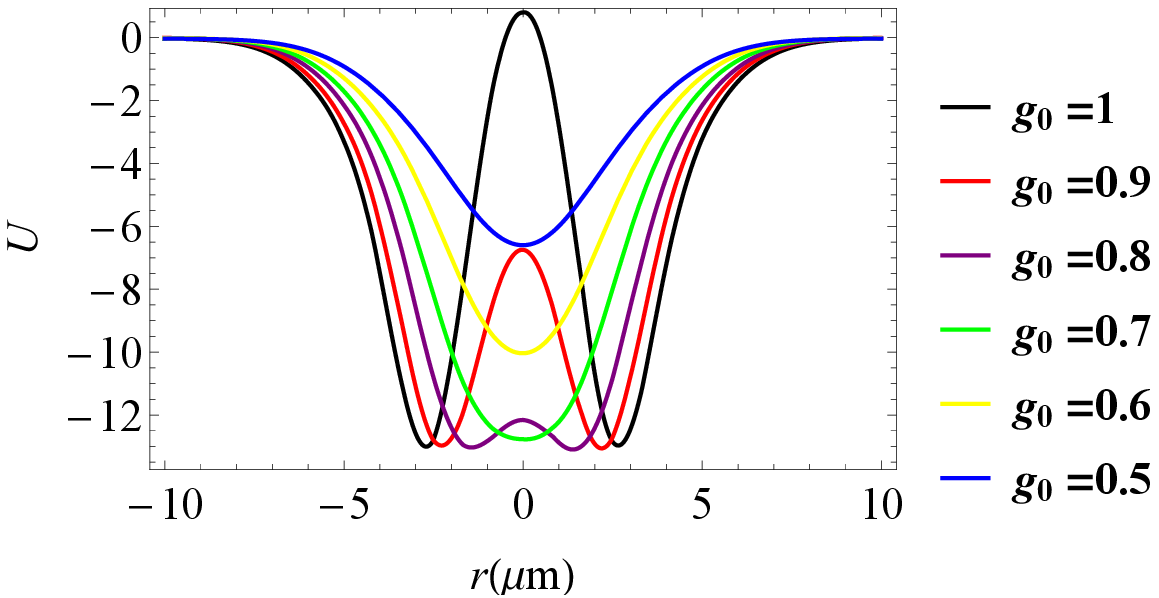}
\vskip 0.5cm \small \rm Fig.\hspace{0.1cm}1. Spatial profile of the OTP when atoms are trapped by fs Gaussian laser pulses under $ w= 5\, \mu m$, $\tilde{\rho}_{11}(-\infty)= 1$ and for different Rabi frequency constants. Other used parameters are $v=100 \, m/s$, $\Delta_{31}=\Delta_{42}=-2.0\,rad/fs$, $\Delta_{32}=0 $ and $\Phi=\pi$.
\vskip 0.8cm \noindent \normalsize
\end{figure}

 The authors of both related papers in this issue \cite{jiang,kumar5} showed that the trap split occurs when the incident laser power or Rabi frequency exceeds a threshold level. Note that trap split in Okamoto's study, is generated when the incident laser power exceeds a threshold level, while in atomic physics, the peak-laser power is related to the so-called Rabi frequency and so we can expect that the laser profile plays a significant role in this emerging phenomenon. In what follows, we will investigate the effect of the laser profile on the OTP. Figure 1 depicts OTP profile subject to the initial population in level $\vert 1 \rangle$ for $\Delta_{31}=\Delta_{42}=-2\,rad/fs$, $\Delta_{32}=0$, $w= 5\, \mu m$, $\Phi=\pi$ and $v=100 \, m/s$. Obviously, it can be seen in figure 1 that one can control the trap size by tuning the Rabi frequencies for Gaussian pulses.
In addition, when the Rabi frequency exceeds a threshold value, trapping potential splits around the center of the trap which means that we have two minima that atoms could occupy these two positions, yet the splitting does appear in the potential for a very limited set of parameters. These results typically emphasize the needs for research into new approaches to readily generate trap split.

As mentioned previously, different controlling methods of the trap split in atomic system via relative phase, frequency and intensity of applied fields were reported. However, to the best of our knowledge, no attempt has yet been made to permanently generate trap split. In the following, we show that the trap split always occurs using the $LG_{0}^{1}$ fs laser pulses. We show this fact in figure 2, which spatial profile of the OTP is plotted when atoms are trapped by such pulses. Other parameters are the same as in figure 1. Trap split always occurs even for small values of the Rabi frequencies; a feature that can be understood intuitively from figure 2.
 
To understand the physics behind this phenomenon, we first point out that the OTP is explicitly dependent on the off-diagonal density matrix elements between relevant transitions: $ (\tilde{\rho}_{ij}+\tilde{\rho}_{ji}), i\in \lbrace1,2\rbrace$ and $j \in \lbrace3,4\rbrace$. For the case of the Gaussian laser pulses, a similar argument to that used in the Okamoto's study \cite{jiang}, will demonstrate the physics of the trap split. When the Rabi frequencies are lower than the threshold, these elements and therefore the OTP have only one minimum, which appears at $r=0$. As the Rabi frequencies exceed a threshold level, on the other hand, the position of the minimal does not appear at the center and the trap position is changed in such a way that splitting can occur. When we use the doughnut pulses, the situation changes; value of the off-diagonal density matrix elements at $r=0$ would be zero. As a result, by multiplying this function by the corresponding Rabi frequencies, trap split is generated. Another explanation relates to spatial probability distribution of the levels, which is given in \cite{kumar5}. It may be mentioned once more that here we present the results for incident nonvortex LG laser pulses, as we do not find any relation between the phase factor and the the trap split. Indeed, by recomputing the figures with the phase factor, the same results is obtained and so, throughout the paper, we evaluated the figures with the inclusion of the phase factor. 

\begin{figure}[t]
\centering
\includegraphics[width=0.7\linewidth]{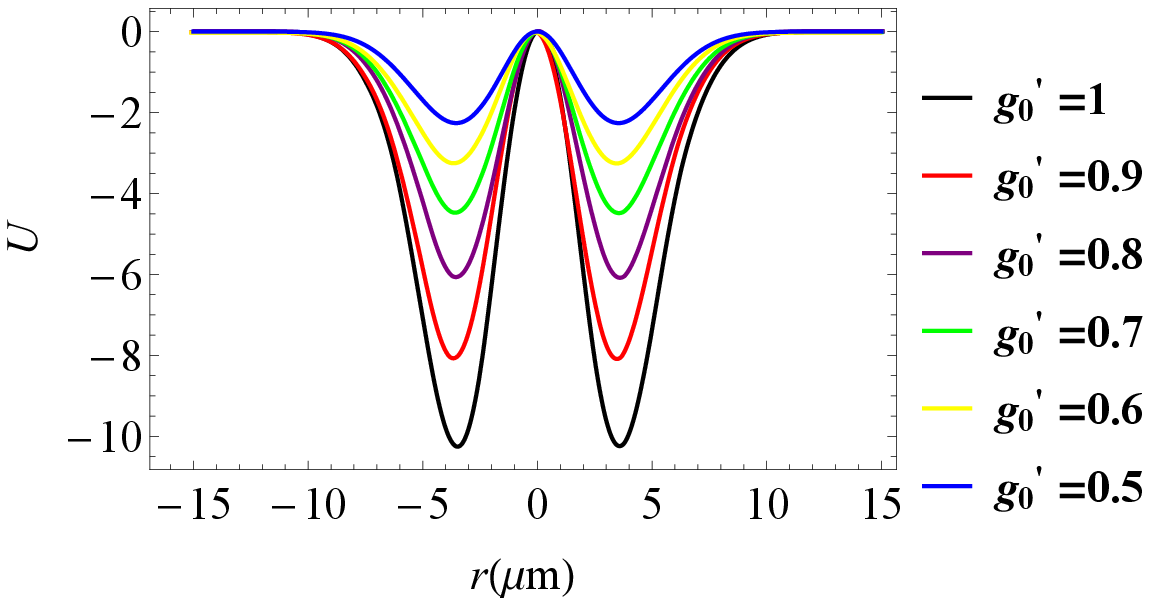}
\vskip 0.5cm \small \rm Fig.\hspace{0.1cm}2. Spatial profile of the OTP when atoms are trapped by fs $LG_{0}^{1}$ laser pulses for different Rabi frequency constants. Other parameters are the same as in figure 1.
\vskip 0.8cm \noindent \normalsize
\end{figure}

As can be observed in figure 1, a prominent feature of traps generated by Gaussian beams is the threshold for the trap split and increasing in the separation between two traps, just above the threshold. For instance, the distance between the traps increases from $4.5\,\mu m$ for $g_{0}=0.9\,rad/fs$ to $5.5\, \mu m$ for $g_{0}=1\,rad/fs$, while the separation between two traps generated by the LG beams remains unchanged ($7.1\, \mu m$ for the Rabi frequency constants, $g^{'}_{0}$, ranging from $0.5$ to $1\,rad/fs$). That is, a significant advantage of the traps produced by LG beams is their insensitivity to fluctuation in intensity of the lasers in such a way that as the Rabi frequencies fluctuate by $\pm 25 \%$ around $g^{'}_{0}=0.75\,rad/fs$, the separation between the traps remains constant at $ 7.1\, \mu m$.

\begin{figure}[t]
\centering
\includegraphics[width=0.7\linewidth]{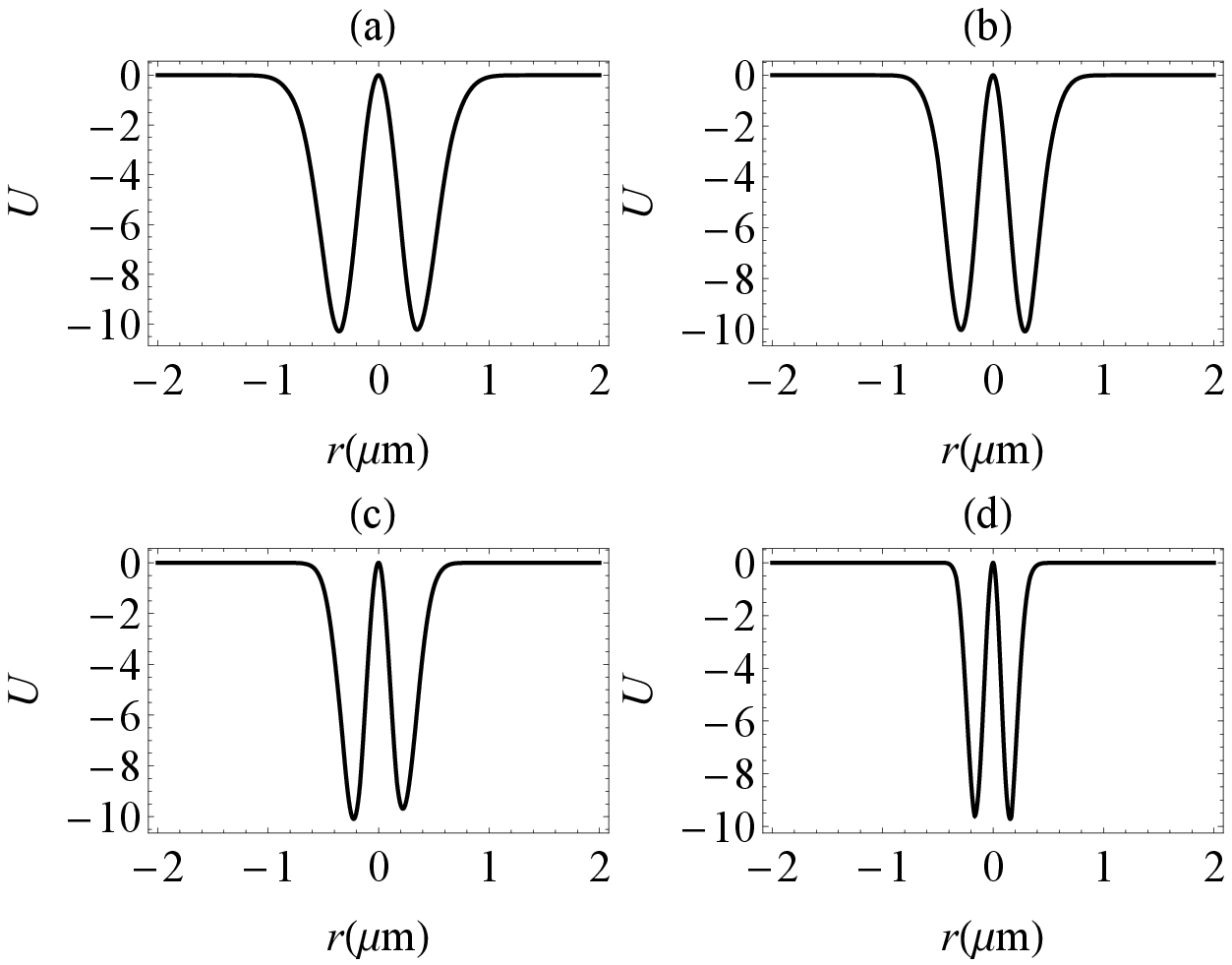}
\vskip 0.5cm \small \rm Fig.\hspace{0.1cm}3. Spatial profile of OTP for $g^{'}_{0}=1\,rad/fs$ and with different beam waists: (a) $w= 0.5\, \mu m$, (b) $w= 0.4\, \mu m$, (c) $w= 0.3\, \mu m$ and (d) $w= 0.2\, \mu m$. Other parameters are the same as in figure 2. For $w= 0.2\, \mu m$, a trap with a size of $157\, nm$ is formed.
\vskip 0.8cm \noindent \normalsize
\end{figure}
We then proceed to investigate the effect of the beam waist on the trap size which is determined by the full width at half-maximum (FWHM). Figure 3 depicts the spatial profile of the OTP under the same parameters as in figure 2 and with four different beam waists: (a) $w= 0.5\, \mu m$, (b) $w= 0.4\, \mu m$, (c) $w= 0.3\, \mu m$ and (d) $w= 0.2\, \mu m$. It can be seen from this figure that the trap size decreases with the decreasing the beam waist in such a way that for $w= 0.2\, \mu m$, two optical traps with the FWHM of $157\, nm$ can be formed, which are about 5 times smaller than the wavelength of light. Unlike some conventional techniques for nanometer-scale optical trap, the narrowing of the trap size is not achieved at the expense of the trap depth; there is no significant reduction in the trap depth even with decreasing the beam waist from $ 5\, \mu m$ to $ 0.2\, \mu m$.

Before going to next section, we should give some comments on a few existing works on Bose gases trapped in an annular geometry. In a recent experiment by Corman \textit{et al} \cite{corman}, supercurrents due to temperature quench has been created by using a quasi-two-dimensional Bose gas trapped in annular geometry. Their experiments are performed with a Bose gas of $ ^{87}$ Rb atoms in which the gas is confined using a harmonic potential along the vertical direction and in the plane, the atoms are trapped in box-like potential. By recording the interference pattern, they have observed a significant fraction of spiral patterns which reveals the presence of a phase winding in the wave function of one of the two clouds. While, our goal is to provide both a convenient and an effective way of generating permanently trap split in a simple scheme in such a way that there is no need to bother with the complexity of their setup, cooling, or obtaining low density. In addition, a prominent feature of traps in our work is the insensitivity to fluctuation in intensity of the lasers. Furthermore, experimental observation of our proposed scheme is easily feasible and promotes better trapping: optical traps with spatial scales smaller than wavelength of light. Then, we proceed to compare Alon's discussion on many-body vortices (MBVs) with our scheme to generate trap split. In their study, two different kinds of the MBVs are discussed, which are made of spatially-partitioned clouds, are fragmented and carry definite total angular momentum\cite{alon}. The MBVs of the first kind are at global minimum of energy, comprising of bosons carrying the same angular momentum. While the second kind are macroscopically fragmented excited states in which macroscopic fractions of bosons carry different angular momentum. By investigating phase diagram based on the solutions of the Gross-Pitaevskii equation in the inner and the outer parts of the trap, Klaiman and Alon showed that one can predict the parameters where many-body vortices occur. Note that the trap split in their study, which density is spread in the disk and the annulus, relies on the main property of the vortices: angular momentum carried by each particle. However, the orbital angular momentum of the LG beams plays no role in our trap split, as we explicitly mentioned in this paper. In other words, trap split in our suggested scheme occurs for incident nonvortex LG laser pulses. Moreover, apart from complexity of the Bose-Einstein condensate setup, an additional difficulty arises in their approach due to fragmentation; as their suggested system develops fragmentation with increasing barrier's height.

Finally, we envision that the first experimental demonstration of our scheme can be implemented with alkali atoms. As an example, we can consider hyperfine energy levels of $ ^{87}$ Rb; two upper levels, i.e., $\vert4\rangle$ and $\vert 3\rangle$ correspond, respectively, to the magnetic sublevels $ \vert 5^{2} P_{3/2}, F^{'}=3 \rangle$ and $ \vert 5^{2} P_{3/2}, F^{'}=1\rangle$, while $\vert 5^{2} S_{1/2}, F=1\rangle$ and $\vert 5^{2} S_{1/2}, F=2\rangle$ are chosen to be the lower states ($\vert1\rangle$ and $\vert 2\rangle $), respectively \cite{steck}. We also note that the required ultrashort LG pulses for the experimental observation of our suggested scheme could be generated by a Ti:sapphire oscillator and two computer-generated diffraction gratings \cite{mary}.

 In summary, we investigated the OTP in a four-level double-$\Lambda$ atomic system in multiphoton resonance condition that closed-loop interaction is created by applying four fs $LG_{0}^{1}$ laser beams. This scheme provides a convenient and effective way to permanently generate trap split which could open a new door for trapping and manipulating the particles. Another advantage of the scheme is its insensitivity to fluctuation in intensity of the lasers in such a way that the separation between the traps remains constant. Moreover, optical traps with spatial scales smaller than wavelength of light can be created.

\bibliographystyle{}

\begin{thebibliography}{9}
\footnotesize
\setlength{\itemsep}{0pt}

\bibitem{Ashkin1}
Ashkin A 1970 \textit{Phys. Rev. Lett.} \textbf{24.4} 156
\bibitem{block}
Block S, Goldstein L and Schnapp B 1990 \textit{Nature} \textbf{348} 348-352
\bibitem{1}
Domokos P and Ritsch H 2003 \textit{J. Opt. Soc. Am. B} \textbf{20.5} 1098-1130
\bibitem{2}
Zhang D, Yuan X, Tjin S and Krishnan S 2004 \textit{Opt. Express} \textbf{12.10} 2220-2230
\bibitem{3}
Zhan Q 2004 \textit{Opt. Express }\textbf{12.15} 3377-3382
\bibitem{4}
Chu S, Bjorkholm J E, Ashkin A and Cable A 2005 \textit{Opt. Lett.} \textbf{30} 1797-1799
\bibitem{5}
Day C 2006 \textit{Phys. Today} \textbf{59.1} 26
\bibitem{6}
Zhao C, Cai Y, Lu X and Eyyubo\u{g}lu H T 2009 \textit{Opt. Express} \textbf{17.3} 1753-1765
\bibitem{jiang}
Jiang Y, Narushima T, and Okamoto H 2010 \textit{Nature Phys.} \textbf{6.12} 1005-1009
\bibitem{9}
Shore B W 2011 \emph{Manipulating quantum structures using laser pulses} (Cambridge University Press).
\bibitem{kemp}
Paul N K and Kemp B A 2016 \textit{J. Opt.} \textbf{18} 085402
\bibitem{Ashkin}
Ashkin A, Dziedzic J M, Bjorkholm J E and Chu S 1986 \textit{Opt. Lett.} \textbf{11.5} 288-290
\bibitem{Li}
Li H, Zhou D, Browne H and Klenerman D 2006 \textit{J. Am. Chem. Soc.} \textbf{128.17} 5711-5717
\bibitem{Tsuboi}
Tsuboi Y, Shoji T, Nishino M, Masuda S, Ishimori K and Kitamura N 2009 \textit{Appl. Surf. Sci.} \textbf{255.24} 9906-9908
\bibitem{liu}
Liu H, Luo A P, Wang F Z, Tang R, Liu M, Luo Z C and Zhang H 2014 \textit{Opt. Lett.} \textbf{39.15} 4591-4594
\bibitem{Baylam}
Baylam I, Cizmeciyan M N, Ozharar S, Polat E O, Kocabas C and Sennaroglu A 2014 \textit{Opt. Lett.} \textbf{39.17} 5180-5183
\bibitem{liu2}
Wang C, Li W, Li L, Hao Q, Zhao J and Zeng H 2016 \textit{J. Opt.} \textbf{18(2)} 025503
\bibitem{agate}
Agate B, Brown C T A, Sibbett W and Dholakia K 2004 \textit{Opt. Express }\textbf{12.13} 3011-3017
\bibitem{pan}
Pan L, Ishikawa A and Tamai N 2007 \textit{Phys. Rev. B} \textbf{75.16} 161305
\bibitem{sanz}
Sanz M, de Nalda R, Marco J F, Izquierdo J G, Banares L and Castillejo M 2010 \textit{J. Phys. Chem. C} \textbf{114.11} 4864-4868
\bibitem{11}
Kumar P and Sarma A K 2012 \textit{Phys. Rev. A} \textbf{86.5} 053414
\bibitem{usman}
Usman A, Chiang W Y and Masuhara H 2012 \textit{ J. Photochem. Photobiol. A. Chem} \textbf{234} 83-90
\bibitem{kumar5}
Chakraborty S and Sarma A K 2015 \textit{J. Opt. Soc. Am. B} \textbf{32.2} 270-274
\bibitem{kazemi}
Kazemi S H and Mahmoudi M 2016 \textit{J. Opt. Soc. Am. B} \textbf{33.3} 479-484
\bibitem{busch}
Busch T and Huyet G 2003 \textit{J. Phys. B} \textbf{36} 2553 
\bibitem{allen}
Allen L, Beijersbergen M W, Spreeuw R J C and Woerdman J P 1992\textit{ Phys. Rev. A} \textbf{45.11} 8185
\bibitem{he}
He H, Heckenberg N R and Rubinsztein-Dunlop H 1995 \textit{J. Mod. Opt.} \textbf{42.1} 217-223
\bibitem{wright}
Wright K C, Leslie L S and Bigelow N P 2008 \textit{Phys. Rev. A} \textbf{77.4} 041601
\bibitem{aksenov}
Aksenov V P, Kolosov V V and Pogutsa C E 2013 \textit{J. Opt.} \textbf{15.4 } 044007
\bibitem{chen2}
Zhang W, Wu Z, Wang J, and Chen L 2016 \textit{Chin. Opt. Lett.} \textbf{14} 110501
\bibitem{Kazemi2}
Kazemi S H and Mahmoudi M 2016 \textit{J. Phys. B: At. Mol. Opt. Phys.} \textbf{49} 245401
\bibitem{otsu}
Otsu T, Ando T, Takiguchi Y, Ohtake Y, Toyoda H and Itoh H 2014 \textit{Sci. Rep.} \textbf{4} 4579
\bibitem{Paterson}
Paterson L, MacDonald M P, Arlt J, Sibbett W, Bryant P E and Dholakia K 2001 \textit{Science} \textbf{292.5518} 912-914
\bibitem{soar}
Soares W C, Moura A L, Canabarro A A, De Lima E and Hickmann J M 2015 \textit{Opt. Lett.} \textbf{40.22} 5129-5131
\bibitem{23}
Korsunsky E A, Leinfellner N, Huss A, Baluschev S and Windholz L 1999 \textit{Phys. Rev. A} \textbf{59.3} 2302
\bibitem{mystre}
Mystre P 2001 \emph{Atom optics} (Springer: Berlin)
\bibitem{aksarma}
Kumar P and Sarma A K 2011 \textit{Phys. Rev. A} \textbf{84.4} 043402
\bibitem{cook}
Cook R J 1979 \textit{Phys. Rev. A} \textbf{20.1} 224
\bibitem{steck}
Steck D A 2015 \textit{Rubidium 87 D line Data} (http://steck.us/alkalidata)
\bibitem{corman}
Corman L \textit{et al} 2014 \textit{Phys. Rev. Lett.} \textbf{113} 135302 
\bibitem{alon}
Klaiman S and Alon O E 2016 \textit{Journal of Physics: Conference Series} \textbf{691.1} 012015 
\bibitem{mary}
Mariyenko I G, Strohaber J and  Uiterwaal C J G J 2005 \textit{Opt. Express} \textbf{13.19} 7599-7608 
\end{thebibliography}

\end{document}